\documentclass[12pt]{article}
\usepackage{latexsym}
\begin{document}

\title{{\bf EXTENDED\\ ELECTRODYNAMICS}:
\\II. Properties and invariant characteristics\\ of the
non-linear vacuum solutions}
\author{{\bf S.Donev}\\
Institute for Nuclear Research and Nuclear Energy,\\
Bulg.Acad.Sci., 1784 Sofia, blvd.Tzarigradsko shausee 72\\
e-mail: sdonev@inrne.acad.bg\\ BULGARIA \and \\
{\bf M.Tashkova}\\
Institute of Organic Chemistry with Center of \\ Phytochemistry,
Bulg.Acad.Sci., 1574 Sofia,\\
Acad. G.Bonchev Str., Bl. 9\\BULGARIA}

\date{}

\maketitle
patt-sol/9710001

\begin{abstract}
This paper aims to consider the general properties of the non-linear
solutions to the vacuum equations of {\it Extended Electrodynamics} 
[1].
The $*$-invariance and the conformal invariance of the equations are
mentioned.
It is also proved that all non-linear solutions have zero invariants:
$\frac 12 F_{\mu\nu}F^{\mu\nu}=\frac 12 (*F)_{\mu\nu}F^{\mu\nu}=0$. 
The three
invariant characteristics of the non-linear solutions: {\it 
amplitude, phase}
and {\it scale factor} are introdiced and discussed.
\end{abstract}

\vskip 0.05cm

\section{General properties of the equations}
Recall from [1] the basic relation of {\it Extended Electrodynamics} 
(EED)
\begin{equation}
\vee(\delta\Omega,*\Omega)=\vee(\Phi,*\pi_1\Omega)+
\vee(\Psi,*\pi_2\Omega).
\end{equation}
An external field is called vacuum with
respect to the $EM$-field $\Omega$ if the right hand side of (1) is 
equal
to zero. Then the left hand side of (1) will also be equal to zero, 
so we
get the equations
\begin{equation}
\vee(\delta\Omega,*\Omega)=0.                           %(2)%
\end{equation}
From this coordinate free compactly written expression we obtain the
following equations for the components of $\Omega$ in the basis 
$(e_1,e_2)$:

\begin{eqnarray}
&(\delta F\wedge *F)\otimes e_1\vee e_1 +(\delta *F)\wedge
**F)\otimes e_2\vee e_2 +\\ \nonumber
&+(\delta F\wedge **F+\delta*F\wedge *F)\otimes e_1\vee e_2=0.    
%(3)%
\end{eqnarray}
So, the field equations, expressed through the operator $\delta$ look 
as
follows
\begin{equation}
\delta F\wedge *F =0,\ \delta *F \wedge **F=0,\          %(4)%
\delta *F\wedge *F-\delta F\wedge F=0.
\end{equation}
These equations, expressed through the operator ${\bf d}$ have the 
form
\begin{equation}
*F\wedge *{\bf d}*F=0,\ F\wedge *{\bf d}F=0,\
F\wedge *{\bf d}*F+*F\wedge *{\bf d}F=0.                  %(5)%
\end{equation}
Using the components $F_{\mu\nu}$, we obtain for the equations (4)
\begin{equation}
F_{\mu\nu}(\delta F)^\nu=0,\ (*F)_{\mu\nu}(\delta*F)^\nu=0,\
F_{\mu\nu}(\delta*F)^\nu+(*F)_{\mu\nu}(\delta F)^\nu=0.       %(6)%
\end{equation}
In the same way, for the equations (5) we get
\begin{equation}
(*F)^{\mu\nu}({\bf d}*F)_{\mu\nu\sigma}=0,
\ F^{\mu\nu}({\bf d}F)_{\mu\nu\sigma}=0,\
(*F)^{\mu\nu}({\bf d}F)_{\mu\nu\sigma}+
F^{\mu\nu}({\bf d}*F)_{\mu\nu\sigma}=0.                       %(7)%
\end{equation}

Now we give the 3-dimensional form of the equations in the same order:
\begin{equation}
B\times\left(rotB-\frac{\partial E}{\partial \xi}\right)-EdivE=0,\
E.\left(rotB-\frac{\partial E}{\partial \xi}\right)=0,         %(8)%
\end{equation}

\begin{equation}
E\times\left(rotE+\frac{\partial B}{\partial \xi}\right)-BdivB=0,\
B.\left(rotE+\frac{\partial B}{\partial \xi}\right)=0,         %(9)%
\end{equation}

\[
\left(rotE+\frac{\partial B}{\partial \xi}\right)\times B+
\left(rotB-\frac{\partial E}{\partial \xi}\right)\times E+
BdivE+EdivB=0,\nonumber
\]
\begin{equation}
B.\left(rotB-\frac{\partial E}{\partial \xi}\right)-
E.\left(rotE+\frac{\partial B}{\partial \xi}\right)=0.          %(10)%
\end{equation}
From the second equations of (8) and (9) the well known Poynting
relation follows
\[
div\left(E\times B\right)+
\frac{\partial}{\partial \xi}\frac{E^2+B^2}{2}=0,
\]
and from the second equation of (10), if $E.B=g(x,y,z)$, we obtain the
known from Maxwell theory relation
\[
B.rotB=E.rotE.
\]
The explicit form of equations (8) and (9) should not make us 
conclude,
that the second (scalar) equations follow from the first (vector) 
equations.
Here is an example: let $divE=0,\ divB=0$ and the time derivatives of 
$E$ and
$B$ are zero. Then the system of equations reduces to
\[
E\times rotE=0,\ B\times rotB=0,\ B.rotE=0,\ E.rotB=0.
\]
As it is seen, the vector equations do not require any connection 
between $E$
and $B$ in this case, therefore, the scalar equations, which require
such connection, can not follow from the vector ones. The third 
equations of
(5) and (6) determine (in equivalent way) the energy-momentum
quantities, transferred from $F$ to $*F$, and reversely, in a unit 4-
volume,
with the expressions, respectively
\[
F_{\mu\nu}(\delta *F)^\nu=-(*F)_{\mu\nu}(\delta F)^\nu,\
F^{\mu\nu}({\bf d}*F)_{\mu\nu\sigma}=
-(*F)^{\mu\nu}({\bf d}F)_{\mu\nu\sigma}.
\]
From these relations it is seen, that the 1-forms $\delta F$ and 
$\delta *F$
{\it play the role of "external" currents} respectively for $*F$ and 
$F$. In
the same spirit we could say, that the energy-momentum quantities
$F_{\mu\nu}(\delta F)^\nu$ and $(*F)_{\mu\nu}(\delta *F)^\nu$,
which $F$ and $*F$ exchange with themselves, are equal to zero. And 
this
corresponds fully to our former statements, concerning the physical 
sense of
the equations for the components of $\Omega$.

From the first two equations of (6) and from the expression
for the divergence $\nabla_\nu Q_\mu^\nu$ of the Maxwell's energy-
momentum
tensor $Q_\mu^\nu$ in CED
\begin{equation}
\nabla_\nu Q_\mu^\nu=
\frac {1}{4\pi}\biggl[F_{\mu\nu}(\delta F)^\nu+
(*F)_{\mu\nu}(\delta *F)^\nu\biggr]                                   
 %(11)%
\end{equation}
it is immediately seen that on the solutions of our
equations (6) this divergence is also zero. In view of this {\it we 
assume
the tensor} $Q_\mu ^\nu$, {\it defined by}
\begin{equation}
Q_\mu^\nu=\frac {1}{4\pi}\biggl[\frac 14 F_{\alpha\beta}
F^{\alpha\beta}\delta_\mu^\nu-F_{\mu\sigma} F^{\nu\sigma}\biggr]=\\
\frac {1}{8\pi}\biggl[-F_{\mu\sigma}F^{\nu\sigma}-
(*F)_{\mu\sigma}(*F)^{\nu\sigma}\biggr].                              
%(12)%
\end{equation}
{\it to be the
energy-momentum tensor in} EED. We shall be interested in finding 
explicit
time-stable solutions of finite type, i.e. $F_{\mu\nu}$ to be finite
functions of the three spatial coordinates, therefore, if it turns 
out that
such solutions really exist, then integral conserved quantities can 
be easily
constructed and computed, making use of the 10 Killing vectors on the
Minkowski space-time.  We recall that in CED such finite and time-
stable
solutions in the whole space are not allowed by the Maxwell's 
equations.

We first note, that in correspondence with the requirement for {\it 
general
covariance}, equations (2), given above and presented in different but
equivalent forms, are written down in coordinate free manner. This
requirement is universal, i.e. it concerns all basic equations of a 
theory
and means simply, that the existence of real objects and the 
occurrence of
real processes {\it can not} depend on the local coordinates used in 
the
theory., i.e. on the convenient for us way to describe the local 
character of
the evolution and structure of the natural objects and processes. Of 
course,
in the various coordinate systems the equations and their solutions 
will look
differently. Namely the covariant character of the equations allows 
to choose
the most appropriate coordinates, reflecting most fully the features 
of
every particular case. A typical example for this is the usage of 
spherical
coordinates in describing spherically symmetric fields. Let's not 
forget
also, that the coordinate-free form of the equations permits an easy 
transfer
of the same physical situation onto manifolds with more complicated 
structure
and nonconstant metric tensor. Shortly speaking, {\it the coordinate 
free
form of the equations in theoretical physics reflects the most 
essential
properties of reality, called shortly objective character of the real
phenomena}.

Since the left hand sides of the equations are linear combinations of 
the
first derivatives of the unknown functions with coefficients, 
depending
linearly on these unknown functions, (6) present a special type 
system of
{\it quasilinear first order partial differential equations}. The 
number of
the unknown functions $F_{\mu\nu}=-F_{\nu\mu}$ is 6, and in general, 
the
number of the equations is $3.4=12$, but the number of the independent
equations depends strongly on if the two invariants
$I_{1}=\frac 12 F_{\mu\nu}F^{\mu\nu}$ and
$I_{2}=\frac 12 (*F)_{\mu\nu}F^{\mu\nu}$  are equal to zero or not 
equal to
zero. If $I_2\neq 0$ then $ det (F_{\mu\nu})\neq 0$ and the first two
equations of (6) are equivalent to $\delta F=0$ and $\delta *F=0$, 
which
automatically eliminates the third equation of (6), i.e. in this case 
our
equations reduce to Maxwell's equations.

It is clearly seen from the (7) form of the equations, that the metric
tensor essentially participates (through the $*$-operator applied to 
2-forms
only) in the equations. If we use the $\delta$-operator, then the
metric participates also through the $*$-operator, applied to 3-
forms, but
this does not lead to more complicated coordinate form of the 
equations. It
worths to note that in nonlinear coordinates the metric tensor will
participate with its derivatives, therefore, the very solutions will 
depend
strongly on the metric tensor chosen. This may cause existence or
non-existence of solutions of a given class, e.g. soliton-like ones. 
In our
framework such additional complications do not appear because of the
opportunity to work in global coordinates with constant metric tensor.

We note 2 important invariance (symmetry) properties of our equations.

\vskip 0.5cm
{\bf Property 1}. {\it The transformation $F\rightarrow *F$ does
not change the system}.

The proof is obvious, in fact, the first two equations interchange, 
and the
third one is kept the same. In terms of $\Omega $ this means that if 
$\Omega$
is a solution, then $*\Omega $ is also a solution, which means, in 
turn, that
equations (2) are equivalent to the equations
\begin{equation}
\vee(\Omega,*{\bf d}\Omega)=0.                                %(13)%
\end{equation}

\vskip 0.05cm
{\bf Property 2}. {\it Under conformal change of the metric the
equations do not change}.

The proof of this property is also obvious and is reduced to the 
notice, that
as it is seen from (7), the $*$-operator participates only with its
reduction to 2-forms, and as it is well known, $*_2$ is conformally
invariant.

\vskip 0.5cm
Summing up the first two equations of (8) and (9) we obtain how the
classical Poynting vector changes in time in our more general 
approach:

\[
\frac {\partial}{\partial \xi}(E\times B)=
EdivE+BdivB-E\times rotE-B\times rotB.
\]

In CED the first and the second terms on the right are missing.

Here is an example of static solutions of (2), which are not 
solutions to
Maxwell's equations.
\[
E=(asin\alpha z,\ acos\alpha z,\ 0),
\  B=(b cos\alpha z,\ -b sin\alpha z,\ 0),
\]
where $a,b$ and $\alpha$ are constants. We obtain
\[
rotE=(a\alpha sin\alpha z,\ a\alpha cos\alpha z,\ 0),\
rotB=(b\alpha cos\alpha z,\ -b\alpha sin\alpha z,\ 0)
\]

Obviously, $E\times rotE=0,\ B\times rotB=0,\ E.rotB=0,\ B.rotE=0.$
For the Poynting vector we get $E\times B =(0,0,-ab)$, and for the 
energy
density \linebreak
$w=\frac12 (a^2+b^2)$. Considered in a finite volume, these solutions
could model some standing waves, but we do not engage ourselves with 
such
interpretations, since we do not accept seriously that static $EM$-
fields may
really exist.

\vskip 0.5cm
\section{General properties of the solutions}
It is quite clear that the solutions of our equations are naturally 
divided
into two classes: {\it linear} and {\it nonlinear}. The first class 
consists
of all solutions to Maxwell's vacuum equations, where the name {\it 
linear}
comes from. These solutions are well known and won't discussed here. 
The
second class, called {\it nonlinear}, includes all the rest 
solutions. This
second class is naturally divided into two subclasses. The first 
subclass
consists of all nonlinear solutions, satisfying the conditions
\begin{equation}
\delta F\neq 0,\ \delta *F\neq 0,                    %(14)%
\end{equation}
and the second subclass consists of those nonlinear solutions, 
satisfying one
of the two couples of conditions:
\[
\delta F=0,\ \delta *F\neq 0 ;\ \delta F\neq 0,\ \delta *F=0 .
\]
Further we assume the conditions (14) fulfilled, i.e. the solutions 
of the
second subclass will be considered as particular cases of the first 
subclass.
Our purpose is to show explicitly, that among the nonlinear solutions 
there
are soliton-like ones, i.e. the components $F_{\mu\nu}$ of which at 
any
moment are {\it finite} functions of the three spatial variables with
{\it connected} support. We are going to study their properties and to
introduce corresponding characteristics. First we shall establish 
some of
their basic features, proving three propositions.

\vskip 0.5cm
{\bf Proposition 1.}\ {\it All nonlinear solutions have zero 
invariants}:
\newline
$$
I_1=\frac 12 F_{\mu\nu}F^{\mu\nu}=0,
\ I_2=\frac 12 (*F)_{\mu\nu}F^{\mu\nu}=2\sqrt{det(F_{\mu\nu})}=0.
$$
\indent{\bf Proof}.\ Recall the field equations in the form (2.8):
\[
F_{\mu\nu}(\delta F)^\nu=0,\ (*F)_{\mu\nu}(\delta*F)^\nu=0,\
F_{\mu\nu}(\delta*F)^\nu+(*F)_{\mu\nu}(\delta F)^\nu=0.
\]
It is clearly seen that the first two groups of these equations may be
considered as two linear homogeneous systems with respect to $\delta 
F^\mu $
and $\delta *F^\mu $ respectively. In view of the nonequalities 
(2.14) these
homogeneous systems have non-zero solutions, which is possible only if
$det(F_{\mu\nu})=det(*F)_{\mu\nu})=0$, i.e. if  $I_2=2E.B=0$. Further,
summing up these three systems of equations, we obtain
\[
(F+*F)_{\mu\nu}(\delta F+\delta *F)^\nu=0.
\]
If now $(\delta F+\delta *F)^\nu\neq 0$, then
\[
0=det(F+*F)_{\mu\nu}=
\left [\frac 12 (F+*F)_{\mu\nu}(*F-F)^{\mu\nu}\right ]^2=
\frac 14\left [-2F_{\mu\nu}F^{\mu\nu}\right ]^2=(I_1)^2.
\]
If $\delta F^\nu=-(\delta *F)^\nu\neq 0$, we sum up the first two 
systems and obtain \linebreak $(*F-F)_{\mu\nu}(\delta *F)^\nu=0$. 
Consequently,
\[
0=det(*F-F)_{\mu\nu}=
\left [\frac 12 (*F-F)_{\mu\nu}(-F-*F)^{\mu\nu}\right]^2 = 
\frac 14\left [2F_{\mu\nu}F^{\mu\nu}\right ]^2=(I_1)^2.
\]
This completes the proof.

Recall that in this case the energy-momentum tensor $Q_{\mu\nu}$ has 
just one
isotropic eigen direction and all other eigen directions are space-
like.
Since all eigen directions of $F_{\mu\nu}$ and $*F_{\mu\nu}$ are eigen
directions of $Q_{\mu\nu}$ too, it is clear that $F_{\mu\nu}$ and
$(*F)_{\mu\nu}$ can not have time-like eigen directions. But the 
first two
systems  of (2.8) require $\delta F$ and $\delta *F$ to be eigen 
vectors of
$F$ and $*F$ respectively, so we obtain
\begin{equation}
(\delta F).(\delta F)\leq 0,\ (\delta *F).(\delta *F)\leq 0.    %(15)%
\end{equation}

\vskip 0.5cm
{\bf Proposition 2.}\ {\it All nonlinear solutions satisfy the 
conditions}
\begin{equation}
(\delta F)_\mu (\delta *F)^\mu =0,\
\left|\delta F\right|=\left|\delta *F\right|             %(16)%
\end{equation}
\indent{\bf Proof}.\  We form the inner product
$i(\delta *F)(\delta F\wedge *F)=0$  and get
\[
(\delta *F)^\mu(\delta F)_\mu(*F)-\delta F\wedge (\delta
*F)^\mu(*F)_{\mu\nu}dx^\nu=0.
\]
Because of the obvious nulification of the second term the first term 
will be
equal to zero (at non-zero $*F$) only if
$(\delta F)_\mu (\delta *F)^\mu=0$.

Further we form the inner product
$i(\delta *F)(\delta F\wedge F-\delta *F\wedge *F)=0$ and obtain

\[
(\delta *F)^\mu(\delta F)_\mu F-
\delta F\wedge (\delta *F)^\mu F_{\mu\nu}dx^\nu-
\]
\[
-(\delta *F)^2(*F)+
\delta *F\wedge(\delta *F)^\mu (*F)_{\mu\nu}dx^\nu=0.
\]
Clearly, the first and the last terms are equal to zero. So, the inner
product by $\delta F$ gives
\[
(\delta F)^2 (\delta *F)^\mu F_{\mu\nu}dx^\nu-
\left[(\delta F)^\mu (\delta *F)^\nu F_{\mu\nu}\right]\delta F
+(\delta *F)^2(\delta F)^\mu (*F)_{\mu\nu}dx^\nu=0.
\]
The second term of this equality is zero. Besides,
$(\delta *F)^\mu F_{\mu\nu}dx^\nu=\linebreak
-(\delta F)^\mu (*F)_{\mu\nu}dx^\nu.$ So,
\[
\left[(\delta F)^2-
(\delta *F)^2\right](\delta F)^\mu(*F)_{\mu\nu}dx^\nu=0.
\]
Now, if $(\delta F)^\mu(*F)_{\mu\nu}dx^\nu\neq 0$, then the relation
$\left|\delta F\right|=\left|\delta *F\right|$ follows immediately.
If $(\delta F)^\mu(*F)_{\mu\nu}dx^\nu=0=-(\delta *F)^\mu 
F_{\mu\nu}dx^\nu$
according to the third equation of (6), we shall show that
 $(\delta F)^2=(\delta *F)^2=0$. In fact, forming the inner product
$i(\delta F)(\delta F\wedge *F)=0$ , we get
\[
(\delta F)^2*F-\delta F\wedge (\delta F)^\mu (*F)_{\mu\nu}dx^\nu=
(\delta F)^2*F=0.
\]
In a similar way, forming the inner product 
$i(\delta *F)\delta *F\wedge F=0$ we have
\[
(\delta *F)^2 F-
\delta (*F)\wedge (\delta *F)^\mu F_{\mu\nu}dx^\nu=
(\delta *F)^2 F=0.
\]
This completes the proof. \newline
We just note that in this last case the isotropic
vectors $\delta F$ and $\delta *F$ are eigen vectors of $Q_{\mu\nu}$ 
too, and
since $Q_{\mu\nu}$ has just one isotropic eigen direction, we 
conclude that
$\delta F$ and $\delta*F$ are colinear.

In order to formulate the third proposition, we recall [2]
that at zero invariants $I_1=I_2=0$ the following representation 
holds:
\[
F=A\wedge \zeta,\ *F=A^*\wedge \zeta,
\]
where $\zeta$ is the only (up to a scalar factor) isotropic eigen 
vector of
$Q_\mu^\nu $ and the relations $A.\zeta=0,\ A^*.\zeta=0$. Having this 
in view
we shall prove the following

{\bf Proposition 3.}\ {\it All nonlinear solutions satisfy the 
relations}
\begin{equation}
\zeta^\mu(\delta F)_\mu=0,\ \zeta^\mu(\delta *F)_\mu=0.       %(17)%
\end{equation}
\indent{\bf Proof}.\ We form the inner product
 $i(\zeta)(\delta F\wedge *F)=0$ :
\[
\left[\zeta^\mu(\delta F)_\mu\right]*F-
\delta F\wedge(\zeta)^\mu(*F)_{\mu\nu}dx^\nu=
\]
\[
=\left[\zeta^\mu (\delta F)_\mu\right]A^*\wedge\zeta-
(\delta F\wedge\zeta)\zeta^\mu (A^*)_\mu
+(\delta F\wedge A^*)\zeta^\mu \zeta_\mu=0.
\]
Since the second and the third terms are equal to zero and $*F\neq 
0$, then
$\zeta^\mu(\delta F)_\mu=0$. Similarly, from the equation
$(\delta *F)\wedge F=0$ we get \linebreak
$\zeta^\mu(\delta *F)_\mu=0$. The proposition is proved.

\vskip 0.5cm
\section{Algebraic properties of the nonlinear \\ solutions}
Since all nonlinear solutions have zero invariants $I_1=I_2=0$ we can 
make a
number of algebraic considerations, which clarify considerably the 
structure
and make easier the study of the properties of these solutions. As we
mentioned earlier, all eigen values if $F$, $*F$ and $Q_{\mu\nu}$ in 
this
case are zero, and the eigen vectors can not be time-like. There is 
only one
isotropic direction, defined by the isotropic vectors $\pm\zeta$ and 
the
representations $F=A\wedge \zeta$,\linebreak $*F=A^*\wedge\zeta$ hold,
moreover, we have  $A.A^*=0,\ A^2=(A^*)^2\leq 0, 
A.\zeta=A^*.\zeta=0$. Recall
that the two 1-forms $A$ and $A^*$ are defined up to isotropic 
additive
factors, colinear to $\zeta$. The above representation of $F$ and $*F$
through $\zeta$ shows that these factors do not contribute to $F$ and 
$*F$,
therefore, we assume further that, {\it these additive factors are 
equal to
zero}.

We express now $Q_{\mu\nu}$ through $A$, $A^*$ and $\zeta$. First we 
normalize
the vector $\zeta$. This is possible, because it is an isotropic 
vector, so
its time-like component $\zeta_4$ is always different from zero. We 
divide
$\zeta_\mu$ by $\zeta_4$ and get the vector
${\bf V}=({\bf V}^1,{\bf V}^2,{\bf V}^3,1)$, defining, of course, the 
same
isotropic direction. Now we make use of the identity,
\begin{equation}
\frac 12 F_{\alpha\beta}G^{\alpha\beta}\delta_\mu^\nu=
F_{\mu\sigma}G^{\nu\sigma}-(*G)_{\mu\sigma}(*F)^{\nu\sigma},      
%(18)%
\end{equation}
where we put
$F_{\mu\nu}$ instead of $G_{\mu\nu}$. Having in view that 
$I_1=\frac 12 F_{\mu\nu}F^{\mu\nu}=0$, we obtain
$F_{\mu\sigma}F^{\nu\sigma}=(*F)_{\mu\sigma}(*F)^{\nu\sigma}$. So, the
energy-momentum tensor looks as follows
\[
Q_\mu^\nu=-\frac{1}{4\pi}F_{\mu\sigma}F^{\nu\sigma}=
-\frac{1}{4\pi}(*F)_{\mu\sigma}(*F)^{\nu\sigma}=
\]
\begin{equation}
=-\frac{1}{4\pi}(A)^2{\bf V}\mu{\bf V}^\nu=
-\frac{1}{4\pi}(A^*)^2{\bf V}_\mu{\bf V}^\nu.               %(19)%
\end{equation}

This choice of  $\zeta={\bf V}$ determines the following energy 
density
\linebreak $4\pi Q_4^4=\left|A\right|^2=\left|A^*\right|^2$.

We consider now the influence of the conservation law $\nabla _\nu
Q_\mu^\nu =0$ on ${\bf V}$.
\[
\nabla_\nu Q_\mu^\nu=-A^2{\bf V}^\nu \nabla_\nu{\bf V}_\mu-
{\bf V}_\mu\nabla_\nu\left(A^2{\bf V}^\nu\right)=0.
\]
This relation holds for every $\mu =1,2,3,4$. We consider it for 
$\mu=4$ 
and get ${\bf V}^\nu\nabla_\nu (1)={\bf V}^\nu{\partial_\nu}(1)=0$.
Therefore,
${\bf V}_4\nabla_\nu\left(A^2 {\bf V}^\nu\right)=
\nabla_\nu\left(A^2 {\bf V}^\nu\right)=0$. Since $A^2\neq 0$, 
we obtain that ${\bf V}$ satisfies the equation
\[
{\bf V}^\nu\nabla_\nu{\bf V}^\mu=0,
\]
which means, that ${\bf V}$ is a {\it geodesic} vector field, i.e. the
integral trajectories of ${\bf V}$ are isotropic geodesics, or
isotropic straight lines. Hence, {\it every nonlinear solution $F$ 
defines
unique isotropic geodesic direction in the Minkowski space-time}. This
important consequence allows a special class of coordinate systems, 
called
further $F$-adapted, to be introduced. These coordinate systems are 
defined
by the requirement, that the trajectories of the unique ${\bf V}$, 
defined by
$F$, to be parallel to the $(z,\xi)$-coordinate plain. In such a 
coordinate
system we have ${\bf V}_\mu=(0,0,\varepsilon,1),\  \varepsilon=\pm1$. 
Further
on, we shall work in such arbitrary chosen but fixed $F$-adapted 
coordinate
system, defined by the corresponding $F$ under consideration.

We write down now the relations $F=A\wedge{\bf V},\ *F=A^*\wedge{\bf 
V}$
component-wise, take into account the values of ${\bf V}_\mu$ in the
$F$-adapted coordinate system and obtain the following explicit 
relations:
\[
F_{12}=F_{34}=0,\ F_{13}=
\varepsilon F_{14},\ F_{23}=\varepsilon F_{24},
\]
\[
(*F)_{12}=(*F)_{34}=0,\ (*F)_{13}=\varepsilon (*F)_{14}=-F_{24},\
(*F)_{23}=\varepsilon (*F)_{24}=F_{14},
\]
\begin{equation}
A=\left(F_{14},F_{24},0,0\right),
\ A^*=\left(-F_{23},F_{13},0,0\right)=
\left(-\varepsilon A_2,\varepsilon A_1,0,0\right).                
%(20)%
\end{equation}
Clearly, the 1-forms $A$ and $-A^*$ can be interpreted as {\it 
electric} and
{\it magnetic} fields respectively. Only 4 of the components 
$Q_\mu^\nu$ are
different from zero, namely: $Q_4^4=-Q_3^3=\varepsilon Q_3^4=
-\varepsilon Q_4^3=\left|A^2\right|$. Introducing the notations
$F_{14}\equiv u,\ F_{24}\equiv p$, we can write
\[
F=\varepsilon udx\wedge dz + udx\wedge d\xi + 
\varepsilon pdy\wedge dz + pdy\wedge d\xi
\]
\[
*F=-pdx\wedge dz + \varepsilon pdx\wedge d\xi + udy\wedge dz +
\varepsilon udy\wedge d\xi.
\]
In the important for us {\it spatially finite } case, i.e. when the 
functions
$u$ and $p$ are finite with respect to the spatial variables 
$(x,y,z)$, for
the integral energy $W$ and momentum ${\bf p}$ we obtain
\[
W=\int{Q_4^4}dxdydz=\int{(u^2+p^2)}dxdydz<\infty,
\]
\begin{equation}
{\bf p}=\left(0,0,\varepsilon \frac Wc\right),\
\rightarrow c^2|{\bf p}|^2-W^2=0.                      %(21)%
\end{equation}

Now we show how the nonlinear solution $F$ defines at every point a
pseudoorthonormal basis in the corresponding tangent and cotangent 
spaces.
The nonzero 1-forms $A$ and $A^*$ are normed to ${\bf A}=A/|A|$ and
${\bf A^*}=A^*/|A^*|$. Two new unit 1-forms ${\bf R}$ and ${\bf S}$ 
are
introduced through the equations:
\[
{\bf R}^2=-1,\ {\bf A}^\nu {\bf R}_\nu=0,\  ({\bf A^*})^\nu {\bf 
R}_\nu=0,\
{\bf V}^\nu{\bf R}_\nu=\varepsilon,\ {\bf S}={\bf V}+\varepsilon{\bf 
R}.
\]
The only solution of the first 4 equations is ${\bf R}_\mu=(0,0,-
1,0)$. Then
for ${\bf S}$ we obtain ${\bf S}_\mu=(0,0,0,1)$. Clearly, ${\bf R}^2=-
1$ and
${\bf S}^2=1$. This pseudoorthonormal (co-tangent) basis
$({\bf A, A^*, R, S})$ is carried over to a
(tangent) pseudoorthonormal basis by means of the pseodometric $\eta$.

We proceed further to introduce the concepts of {\it amplitude} and 
{\it
phase} in a coordinate-free manner. First, of course, we look at the
invariants, we have: $I_1=I_2=0$. But in our case we have got another
invariant, namely, the module of the 1-forms $A$ and $A^*$: 
$|A|=|A^*|$.
Let's begin with the {\it amplitude}, which shall be denoted by 
$\phi$. As
it's seen from the above obtained expressions, the magnitude $|A|$ of 
$A$
coincides with the square root of the energy density in any $F$-
adapted
coordinate system. As we noted before, this is the sense of the 
quantity
amplitude. So, we define it by the module of $|A|=|A^*|$. We give now 
two
more coordinate-free ways to define the amplitude.

Recall first, that at every point, where the field is different from 
zero, we
have three bases: the pseudoopthonormal coordinate basis 
$(dx,dy,dz,d\xi)$,
the pseudoorthonormal basis $\chi^0=({\bf A},\varepsilon{\bf 
A^*},{\bf R},
{\bf S})$ and the pseudoorthogonal basis $\chi=(A,\varepsilon 
A^*,{\bf R},
{\bf S})$. The matrix $\chi_{\mu\nu}$ of  $\chi$ with respect to the
coordinate basis is
\[
\chi_{\mu\nu}=\left\|\matrix{
u  &-p  & 0  &0 \cr
p  & u  & 0  &0 \cr
0  & 0  &-1  &0 \cr
0  & 0  & 0  &1
\cr}\right\|.
\]
We define now the amplitude $\phi$ of the field by
\begin{equation}
\phi=\sqrt{|det(\chi_{\mu\nu})|}.                            %(22)%
\end{equation}

We consider now the matrix ${\cal R}$ of 2-forms
\[
{\cal R}=\left\|\matrix{
udx\wedge d\xi  &-pdx\wedge d\xi  & 0            &0 \cr
pdy\wedge d\xi  & udy\wedge d\xi  & 0            &0 \cr
0               & 0               &-dy\wedge dz  &0 \cr
0               & 0               & 0            &dz\wedge d\xi
\cr}\right\|,
\]
or, equivalently:
\[
{\cal R}=udx\wedge d\xi \otimes(dx\otimes dx)-pdx\wedge d\xi
\otimes(dx\otimes dy) + pdx\wedge d\xi \otimes(dy\otimes dx)+
\]
\[
+udy\wedge d\xi \otimes(dy\otimes dy)-
dy\wedge dz \otimes(dz\otimes dz)+
dz\wedge d\xi \otimes(d\xi\otimes d\xi).
\]
Now we can write
\[
\phi=\sqrt{\frac12 \left|R_{\mu\nu\alpha\beta}
R^{\mu\nu\alpha\beta}\right|}.
\]

We proceed further to define the {\it phase} of the nonlinear 
solution $F$.
We shall need the matrix $\chi^0_{\mu\nu}$ of the basis $\chi^0$ with 
respect
to the coordinate basis. We obtain
\[
\chi^0_{\mu\nu}=\left\|\matrix{
\frac{u}{\sqrt{u^2+p^2}}  &\frac{-p}{\sqrt{u^2+p^2}}  & 0  &0 \cr
\frac{p}{\sqrt{u^2+p^2}}  &\frac{u}{\sqrt{u^2+p^2}}   & 0  &0 \cr
0                         & 0                         &-1  &0 \cr
0                         & 0                         & 0  &1
\cr}\right\|.
\]
The {\it trace} of this matrix is
$$
tr(\chi^0_{\mu\nu})=\frac{2u}{\sqrt{u^2+p^2}}.
$$
Obviously, the inequality $|\frac 12 tr(\chi^0_{\mu\nu})|\leq 1$ is
fulfilled. Now, by definition, the quantity
$\varphi=\frac 12 tr(\chi^0_{\mu\nu})$ will be called {\it phase 
function} of
the solution, and the quantity
\begin{equation}
\theta=arccos(\varphi)=
arccos\left(\frac 12 tr(\chi^0_{\mu\nu})\right)
\end{equation}                                                 %(23)%
will be called {\it phase} of the solution.

Making use of the amplitude $\phi$ and the phase function $\varphi $ 
we can
write
\begin{equation}
u=\phi.\varphi,\ p=\phi.\sqrt{1-\varphi^2}.                     %(24)%
\end{equation}

We note that the couple of 1-forms $A=udx+pdy,\ A^*=-pdx+udy $ 
defines a
completely integrable Pfaff system, i.e. the following equations hold:
\[
{\bf d}A\wedge A\wedge A^* =0, \ {\bf d}A^*\wedge A\wedge A^* =0.
\]
In fact, $A\wedge A^*=(u^2+p^2)dx\wedge dy$, and in every term of
${\bf d}A$ and ${\bf d}A^*$ at least one of the basis vectors $dx$ 
and $dy$
will participate, so the above exteriour products will vanish.

\noindent{\it Remark}. These considerations stay in force also for 
those
linear solutions, which have zero invariants $I_1=I_2=0$. But 
Maxwell's
equations require $u$ and $p$ to be {\it running waves}, so the 
corresponding
phase functions will be also running waves. As we'll see further, the 
phase
functions for nonlinear solutions are arbitrary bounded functions.

We proceed further to define the new and important concept of {\it 
scale
factor} $L$ for a given nonlinear solution. It is defined by
\begin{equation}
L=\frac{|A|}{|\delta F|}=\frac{|A^*|}{|\delta*F|}.              %(25)%
\end{equation}
Clearly, $L$ can not be defined for the linear solutions, 
and in this sense it
is {\it new} and we shall see that it is really {\it important}.

From the corresponding
expressions $F=A\wedge {\bf V}$ and $*F=A^*\wedge {\bf V}$ it
follows that the physical dimension of $A$ and $A^*$ is the same as 
that of
$F$. We conclude that the physical dimension of $L$ coincides with the
dimension of the coordinates, i.e. $[L]=length$. From the definition 
it is
seen that $L$ is an {\it invariant} quantity, and depends on the 
point, in
general. The invariance of $L$ allows to define a time-like 1-form (or
vector field) $f(L){\bf S}$, where $f$ is some real function of $L$. 
So,
every nonlinear solution determines a time-like vector field on $M$.

If the scale factor $L$, defined by the nonlinear solution $F$, is a
{\it finite} and {\it constant} quantity, we can introduce a {\it
characteristic} finite time-interval $T(F)$ by the relation
$$
cT(F)=L(F),
$$
as well, as corresponding {\it characteristic frequency} by
$$
\nu(F)=1/T(F).
$$
In these "wave" terms the scale factor $L$ acquires the sense of "wave
length", but this interpretation is arbitrary and we shall 
not make use of it.

It is clear, that the subclass of nonlinear solutions, which define 
constant
scale factors, factors over the admissible values of the invariant 
$f(L)$.
This makes possible to compare with the experiment. For example, at 
constant
scale factor $L$ if we choose $f(L)=L/c$, then the scalar product of
$(L/c){\bf S}$ with the integral energy-momentum vector, which in the
$F$-adapted coordinate system is $(0,0,\varepsilon W,W)$, gives the 
invariant
quantity $W.T$, having the physical dimension of action, and its 
numerical
value could be easily measured.

\newpage
{\bf References}:\\

[1]. Donev,S., Tashkova,M., {\bf Extended Electrodynamics}: I. {\it 
Basic\\
Notions, Principles and Equations}, submitted for publication.

[2]. Synge,J., {\it Relativity: the Special Theory}, North-Holland
Publ.Comp., Amsterdam, 1958

\end{document}